\documentclass[sigconf]{acmart}

\usepackage{siunitx}
\usepackage{lipsum}
\newcommand\blfootnote[1]{%
	\begingroup
	\renewcommand\thefootnote{}\footnote{#1}%
	\addtocounter{footnote}{-1}%
	\endgroup
}
\usepackage{tabularx}
\usepackage[font=small,skip=10pt]{caption}
\usepackage{booktabs} 
\usepackage{mathrsfs}
\usepackage{mathtools}
\usepackage{todonotes}
\usepackage{siunitx}

\graphicspath{/artwork/}

\DeclareMathAlphabet\mathbfcal{OMS}{cmsy}{b}{n}
\DeclareMathAlphabet\mathcal{OMS}{cmsy}{c}{n}

\setcopyright{rightsretained}

\begin{document}
\title{Maximizing Information Gain for the Characterization of Biomolecular Circuits}

\author{Tim Prangemeier*, Christian Wildner*, Maleen Hanst, and Heinz Koeppl}
\orcid{1234-5678-9012}
\affiliation{%
  \institution{Department of Electrical Engineering and Information Technology,}
\city{Technische Universit\"at Darmstadt}
  \country{Germany}
}
\email{tim.prangemeier;christian.wildner; maleen.hanst; heinz.koeppl@bcs.tu-darmstadt.de}

\renewcommand{\shortauthors}{T. Prangemeier*, C. Wildner*, M. Hanst and H. Koeppl}

\begin{abstract}
Quantitatively predictive models of biomolecular circuits are important tools for the design of synthetic biology and molecular communication circuits. The information content of typical time-lapse single-cell data for the inference of kinetic parameters is not only limited by measurement uncertainty and intrinsic stochasticity, but also by the employed perturbations. Novel microfluidic devices enable the synthesis of temporal chemical concentration profiles. The informativeness of a perturbation can be quantified based on mutual information. We propose an approximate method to perform optimal experimental design of such perturbation profiles. To estimate the mutual information we perform a multivariate log-normal approximation of the joint distribution over parameters and observations and scan the design space using Metropolis-Hastings sampling. The method is demonstrated by finding optimal perturbation sequences for synthetic case studies on a gene expression model with varying reporter characteristics.


\end{abstract}

%
%
%


\copyrightyear{2018} 
\acmYear{2018} 
\setcopyright{acmlicensed}
\acmConference[NANOCOM '18]{NANOCOM '18: ACM The Fifth Annual International Conference on Nanoscale Computing and Communication}{September 5--7, 2018}{Reykjavik, Iceland}
\acmBooktitle{NANOCOM '18: NANOCOM '18: ACM The Fifth Annual International Conference on Nanoscale Computing and Communication, September 5--7, 2018, Reykjavik, Iceland}
\acmPrice{15.00}
\acmDOI{10.1145/3233188.3233217}
\acmISBN{978-1-4503-5711-1/18/09}


\maketitle
\section{Introduction}

Synthetic biology enables the programming of living cells on the molecular level  and promises to be the basis of a range of new technologies. \blfootnote{* Both authors contributed equally.} Quantitatively predictive models of the involved biomolecular processes facilitate the design of synthetic parts and molecular communications circuits. Assessment of the kinetic model parameters is often characterized by a large degree of uncertainty, not only due to measurement uncertainty but also inherent cell-heterogeneity and intrinsic stochasticity of the involved process \cite{Elowitz2002}. Continuous-time Markov chains provide a reasonable approximation of the dynamics of stochastic reaction networks \cite{McQuarrie1967}, such as biomolecular circuits. 

\begin{figure}[h]
	\captionsetup{width=0.8\linewidth}
	\includegraphics[width =0.7\linewidth]{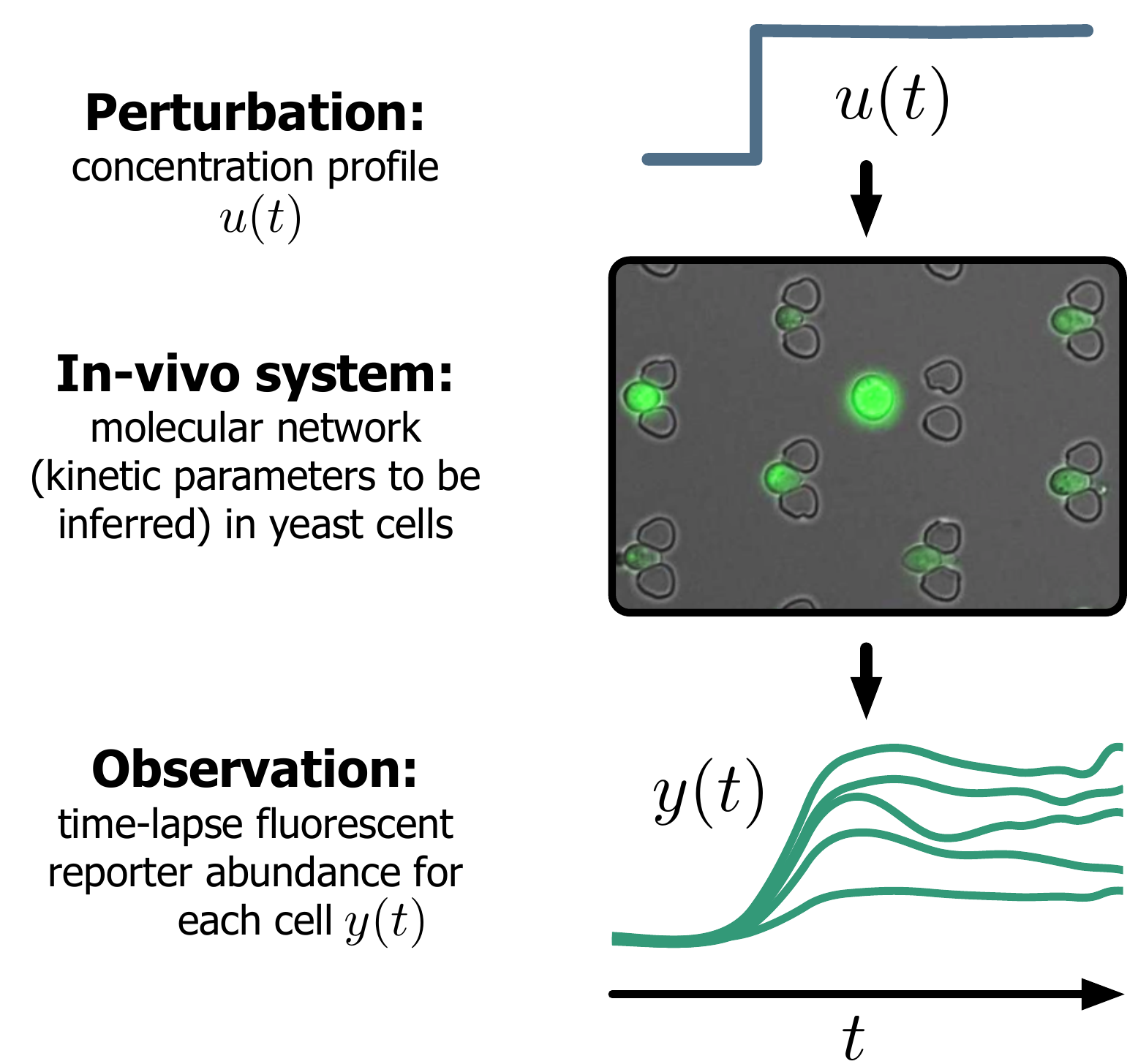}
	\caption{Schematic of the time-lapse single-cell experiments upon which parameter inference for the biomolecular circuit models considered here is based.}
	\label{fig:introGraphic}
\end{figure}

The experimental process of perturbing and observing cells for the purpose of parameter inference is shown schematically in Fig. \ref{fig:introGraphic}. Time-lapse fluorescence readouts $y(t)$ of single-cells contain information about the temporal behaviour of biomolecular circuits upon some chemical perturbation $u(t)$ \cite{Crane2014, Schneider2017}. Bayesian inference can for example be employed to calibrate model parameters \cite{Zechner2014}. The information gained is not only limited by experimental constraints, measurement uncertainty and inherent heterogeneity within populations of clonal cells, but also by the dynamics of the perturbation. 

In the past, technical constraints have limited the employed perturbations to a single step or pulse. Recent advances in microfluidics enable the generation of well defined temporal concentration profiles \cite{Ainla2009, Unger2011}. In this study, we exploit this new capability and find approximately optimal perturbation profiles $u(t)$ for the inference of biomolecular model parameters. 

Bayesian optimal experiment design for the inference of stochastic reaction networks was studied theoretically in \cite{Nandy2012} for the case of complete observations. These considerations were extended to incomplete and noisy observations in \cite{Zechner2012a}, where informativeness was quantified using the expected log-determinant of the posterior covariance matrix. A Monte Carlo gradient descent method was used to approximately solve the resulting optimization problem.

Other approaches focus on the Fisher information matrix as a measure of informativeness. In \cite{Komorowski2011} the linear noise approximation is used to obtain a differential equation for the Fisher information matrix.  The method proposed in \cite{Ruess2013,Ruess2015} combines a  Gaussian approximation of the Fisher information with moment-based inference \cite{Zechner2012}.

Here, we follow the Bayesian approach and use the mutual information as a measure of informativeness. In the context of cell biology, such an approach is discussed in \cite{Liepe2013}. Based on an ordinary differential equation model of the chemical kinetics, mutual information was estimated using Monte Carlo simulations. 



\subsection{Stochastic Chemical Reaction Kinetics} \label{sec:chemical_kinetics}
The observed behaviour of fluorescent  reporter proteins stems from the interaction of various biomolecules in a system of chemical reactions. We consider the dynamics of such a system as a continuous-time Markov chain (CTMC) $\mathbf{X}(t)$ on the time interval $[0,T]$, assuming the system is well mixed. 

The species $\mathcal{X}_1,...,\mathcal{X}_n$ interact in $\nu$ reactions. Each reaction is associated with a stoichiometric change vector $\mathbf v_i$ encoding the net change of the molecule abundance and a propensity function $h_i(\mathbf{x} ) = c_i g_i( \mathbf x )$. Here, $c_i$ is the kinetic rate constant and $g_i(\mathbf x)$ is a possibly nonlinear function of the state. In this study, we consider mass-action kinetics where $g_i(\mathbf x)$ essentially corresponds to the product of the involved reactant abundances. To simplify notation, we also introduce the vector of the rate constants $\mathbf c = (c_1, \ldots, c_v)^T$. 

Intuitively, $h_i(\mathbf x) dt$ is the probability that reaction $i$ fires in the infinitesimal time interval $dt$. Consequently, the combined propensity of any reaction occurring follows as $h_0(\mathbf{x}) =\sum_{i=1}^{\nu} h_i(\mathbf{x})$ and the waiting time between  consecutive reaction events is exponentially distributed. In particular, if $\tau_j$ denotes the time of the $j$-th reaction event, the waiting time is given by
$$((\tau_j - \tau_{j-1}) \mid \mathbf{X}(\tau_{j-1}) = \mathbf{x} ) \sim \mathrm{Exp}(h_0(\mathbf{x} ).$$ 
The probability that the $i$-th reaction has caused the event at $\tau_j$ is given by
\begin{equation*}
p( \mathbf X(\tau_j) = \mathbf x + \mathbf v_i \mid X(\tau_{j-1} ) = \mathbf x , \tau_j-\tau_{j-1}  ) = \frac {h_i(\mathbf{x} )}{h_0(\mathbf{x} )} \, .
\end{equation*}
The above equations lie at the heart of the stochastic simmulation algorithm that allows to generate statistically exact sample paths of the process \cite{Anderson2007}. More information on stochastic modelling of chemical reactions systems  can be found in \cite{Wilkinson2012}. 

In practice, we cannot measure full sample paths. Instead, one typically obtains noisy readouts at the sample times $t_1, \ldots, t_N$. The measurement $Y(t_n) = y_n$ is connected to the latent state $\mathbf X(t_n)$ by the observation model $p( y_n \mid \mathbf x(t_n) )$. We combine all measurements into a vector $\mathbf y = (y_1, \ldots, y_N)^T$.

The central goal of model calibration is to infer the kinetic rate parameters $\mathbf c$ from the noisy measurements $\mathbf y$. We adopt a Bayesian approach and choose a prior distribution $p(\mathbf c)$ over the model parameters. This prior distribution reflects the uncertainty about the model parameters before the data $\mathbf y$ is obtained.  Assuming that the measurements are independent given the latent path, the joint density over all quantities becomes
\begin{equation*}
p( \mathbf c , \mathbf x , \mathbf y ) =  p( \mathbf c) p( \mathbf x \mid \mathbf c) \prod_{n=1}^N p( y_n \mid \mathbf x(t_n) ) \, .
\end{equation*}
where $p(\mathbf c)$ refers to prior distribution over the parameters and $\mathbf x$ denotes the full latent sample path. 

In Bayesian inference, the quantity of interest is the parameter posterior 
\begin{equation*}
p( \mathbf c \mid \mathbf y) = \frac{\int  p( \mathbf c , \mathbf x , \mathbf y )  \, d \mathbf x }{ p( \mathbf y ) } \, .
\end{equation*}
This distribution is intractable because it requires the evaluation of infinite dimensional integrals with respect to the latent paths $\mathbf X$. 

Several sampling-based approaches to approximate the above posterior have been developed recently \cite{Zechner2014, Golightly2011}. Since these approaches are computationally intensive, here we focus on an optimal design approach that circumvents the posterior altogether. 

As a specific example that will also serve as the basis for several case studies, consider the following gene expression model 
\begin{equation*}
\begin{aligned}
\mathrm{G^{off}} & \xrightleftharpoons[c_2]{c_1 u(t) } \mathrm{G^{on}}  \quad & \quad \text{mRNA} &\xrightharpoonup{\hspace{2mm}c_5 \hspace{2mm}} \text{mRNA} + \mathrm{P}^{0} \\
\mathrm{G^{on}} &\xrightharpoonup{\hspace{2mm}c_3 \hspace{2mm}} \mathrm{G^{on}} +  \text{mRNA}   \quad & \quad \mathrm{P}^{0} &\xrightharpoonup{\hspace{2mm}c_6 \hspace{2mm}} \mathrm{P}^{1}  \\
\text{mRNA} &\xrightharpoonup{\hspace{2mm}c_4 \hspace{2mm}} \emptyset  \quad & \quad  \mathrm{P}^{1} &\xrightharpoonup{\hspace{2mm}c_7 \hspace{2mm}} \emptyset \, .
\end{aligned}
\end{equation*}
Here, transcription is regulated by a single gene $\mathrm{G}$ that can switch between on and off states. As is common in synthetic biology, the transition from the inactive to the active gene state is modulated by an external perturbation $u(t)$. The input $u(t)$ corresponds to the concentration of the inducer molecule in the cell.  

The mRNA is transcribed from the gene and serves as a template to produce an inactive protein $\mathrm{P}^0$. An additional conversion is necessary to obtain the active reporter protein $\mathrm{P}^1$. This can, for example, reflect maturation or folding times and enables us to consider the properties of various reporter dynamics.

\subsection{Optimal Perturbation Design}
\begin{figure}[h]
	\captionsetup{width=\linewidth,skip=4pt}
	\includegraphics[width = \linewidth]{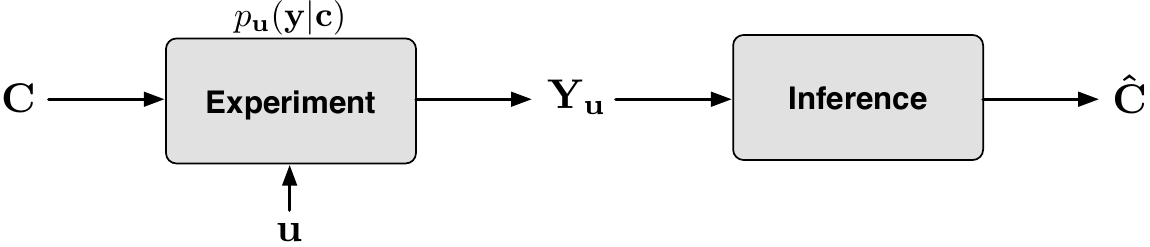}

	\caption{Schematic depiction of the channel capacity approach to optimal experimental design. }
	\label{img:inference_channel}
\end{figure}

The outcome $\mathbf Y$ of an experiment depends on the unknown model parameters $\mathbf C$ and  the perturbation $u(t)$.  We express this in terms of a probabilistic model $p_{u}( \boldsymbol y \mid \mathbf c)$. As illustrated in Fig. \ref{img:inference_channel}, we consider the inference task as an abstract communication problem. In this picture, the probabilistic model $p_{u} ( \boldsymbol y \mid \mathbf c ) $ corresponds to a noisy channel with input $\mathbf C$, where the unknown rate constant chosen by nature from the prior distribution $p(\mathbf c)$. The inference step corresponds to decoding in the communication picture. 

 From an information theoretic point of view, the mutual information $I_u( \mathbf C; \mathbf Y)$  quantifies the average information about $\mathbf C$ contained in the noisy measurement $\mathbf Y$. Hence, a principled objective function for parameter inference is given by
\begin{equation} \label{eqn:objective_function_mi}
J(u) = I_u( \mathbf C; \mathbf Y) = \int p_u(\mathbf c, \mathbf y) \log \frac{  p_u(\mathbf c, \mathbf y)  }{ p(\mathbf c) p_u (\mathbf y) } d \mathbf c d \mathbf y
\end{equation}
and the optimal design choice $u^*$ is given by $u^* = \arg \max_{ u} J(u)$ \cite{Liepe2013}. The quantity $J^* = \max_{ u } I( \mathbf C ; \mathbf Y_{ u} )$  resembles the capacity of a  noisy channel. However, to compute the channel capacity, one maximizes with respect to the input distribution for a fixed channel model, whereas in optimal design, the roles are reversed. 
In the communication picture, we may see $J^*$ as the inference capacity of the experiment $p_{ u} ( \mathbf y \mid  \mathbf c )$ and $u^*$  as the capacity achieving design. 

 In the classical approach to Bayesian experimental design (\cite{Lindley1956}) the objective function  $J( u )$ is given as
\begin{equation} \label{eqn:objective_function_kl}
\begin{split}
J( u) &= \int  p_{ u}(\mathbf y) \left[ \int p_{ u} ( \mathbf c \mid \mathbf y) \log \frac{ p_{ u} ( \mathbf c \mid \mathbf y)}{ p( \mathbf c ) } d \mathbf c \right] d \mathbf y \\
&= \mathbb E_{\mathbf Y_{\mathbf u}} \left[ D_{ KL } [ p_u ( \mathbf c \mid \mathbf y) \, || \, p(\mathbf c  )] \right] \, ,
\end{split}
\end{equation}
where the Kullback-Leibler divergence between posterior and prior distribution corresponds to the reduction in uncertainty (or gain in information) by the particular observation $\mathbf Y = \mathbf y$. Since the outcome $\mathbf Y$ is unknown a priori, the expectation must be taken with respect to all possible outcomes.  It is straightforward to show that Eqn. \ref{eqn:objective_function_kl} and \ref{eqn:objective_function_mi} are equivalent. 

	\vspace{-4mm}


\section{Methods}

Approximations to optimal design are often based on Eqn. \ref{eqn:objective_function_kl} and target the posterior $p(\mathbf c \mid \mathbf y)$ \cite{Chaloner1995}. Here, we take an approach that focuses on  Eqn. \ref{eqn:objective_function_mi} and approximates the joint $p(\mathbf c , \mathbf y)$.

\subsection{Multivariate Log-Normal Approximation}
The idea behind this approximation is to choose a suitable parametric family $p_{\boldsymbol \phi} ( \mathbf c, \mathbf y)$ and find the closest representation of the joint distribution by minimizing Kullback-Leibler divergence with respect to the second argument
\begin{equation} \label{eqn:kl_approximation}
\hat{ \boldsymbol \phi } = \arg \min_{\boldsymbol \phi} D_{KL} [ p( \mathbf c, \mathbf y) \, || p_{\boldsymbol \phi} (\mathbf c, \mathbf y) ]  \, .
\end{equation}
This approach has favourable information theoretic properties in the context of belief approximation \cite{Leike2017}.  

In the case of a stochastic chemical reaction system, a suitable choice of the prior over the rate constants is provided by a product Gamma distribution \cite{Zechner2014}. As an approximate model we choose a multivariate log-normal distribution. Let $\mathbf z = (\mathbf c, \mathbf y)$ denote the joint vector of parameters and observations with $M= \nu+N$ components. Then, the log-normal model corresponds to the density function 
\begin{equation*}
p_{\boldsymbol \mu, \boldsymbol \Sigma} (\mathbf z) =   |2 \pi \boldsymbol \Sigma |^{-\frac{1}{2}} \prod_{k=1}^M z_k^{-1} \exp \left[ -\frac{1}{2} \left( \log ( \mathbf z) - \boldsymbol \mu \right)^T \boldsymbol \Sigma^{-1}  \left( \log ( \mathbf z) - \boldsymbol \mu \right) \right]
\end{equation*}
where the logarithms act component-wise on the vector $\mathbf z$. Here, $\boldsymbol \mu \in \mathbb R^M$ and $\boldsymbol \Sigma \in \mathbb R^{M \times M} $ is a positive semidefinite matrix.

The multivariate log-normal distribution has several desirable properties in the present context. First, it is restricted to positive values. Second it has heavier tails than the multivariate normal distribution. In contrast to a product Gamma distribution it also allows to model correlations between variables.

Setting $\boldsymbol \phi = (\boldsymbol \mu, \boldsymbol \Sigma )$, a short calculation shows that the optimal parameters in the sense of Eqn. \ref{eqn:kl_approximation} are given by
\begin{align*}
\hat{ \boldsymbol \mu } &= \mathbb E \left[ \log \mathbf z \right] \, ,\\
\hat{ \boldsymbol \Sigma } &=  \mathbb E \left[ \left( \log \mathbf z - \hat{ \boldsymbol \mu } \right)  \left( \log \mathbf z - \hat{ \boldsymbol \mu } \right)^T \right] \, .
\end{align*}
Though the expectation with respect to the target distribution $ p( \mathbf c, \mathbf y)$ cannot be evaluated analytically, it is straightforward to estimate the results using samples from the distribution $p(\mathbf c, \mathbf y)$. 

Since the perturbation $u(t)$ modulates one of the rate constants, a stochastic simulation algorithm capable of dealing with time-dependent propensities has to be used \cite{Anderson2007}. Alternatively, one can introduce a dummy species with high firing rate ensuring regular updates of the propensity in segments where $u(t)$ is small. We follow the latter approach here. 

The log-normal distribution permits a closed form solution of the mutual information \cite{Kvalseth1982}. In particular, if $p(\mathbf c, \mathbf y)$ is multivariate log-normal with parameters
\begin{equation*}
\boldsymbol \Sigma = \begin{pmatrix}
\boldsymbol \Sigma_{\boldsymbol C \boldsymbol C } & \boldsymbol \Sigma_{\boldsymbol C \boldsymbol Y }  \\
\boldsymbol \Sigma_{\boldsymbol Y \boldsymbol C } & \boldsymbol \Sigma_{\boldsymbol Y \boldsymbol Y }  \\
\end{pmatrix} \, ,
\end{equation*} 
the mutual information is given by
\begin{equation} \label{eq:kl_ln_approx}
I ( \mathbf C; \mathbf Y ) = \frac{1}{2} \log \left( \frac{ | \boldsymbol \Sigma_{\boldsymbol C \boldsymbol C } | | \boldsymbol \Sigma_{\boldsymbol Y \boldsymbol Y } | }{ | \boldsymbol \Sigma  |} \right) \, .
\end{equation}
Since mutual information is invariant under one-to-one transformations \cite{Kraskov2004}, we would obtain the same result by first applying a log transformation on $\mathbf C$ and $\mathbf Y$ and then performing a Gaussian approximation in the log domain.

The mutual information above may be rewritten as 
\begin{equation*}
I ( \mathbf C; \mathbf Y ) = \frac{1}{2} \log \left( \frac{ | \boldsymbol \Sigma_{\boldsymbol C \boldsymbol C } | }{  |(\boldsymbol \Sigma_{\mathbf C \mid \mathbf Y } | } \right),
\end{equation*}
where $\Sigma_{\mathbf C \mid \mathbf Y} = \boldsymbol \Sigma_{\mathbf C \mathbf C} - \boldsymbol \Sigma_{\mathbf C \mathbf Y} \boldsymbol \Sigma_{\mathbf Y \mathbf Y}^{-1} \boldsymbol \Sigma_{\mathbf Y \mathbf C}$. This implies that changes in the posterior variance are exponentially related to the mutual information. 

Note, that since $\Sigma_{\mathbf C \mathbf C }$ is independent of $u$, the  approximate objective function $J( u)$ is equivalent to the classical D-optimality criterion in the log domain  \cite{Fedorov1997}. 

\subsection{Input Segment Discretization}
In our setup, the input $u(t)$ modulates the activation rate of the gene. Experimentally, the perturbation $u(t)$ is created by varying the concentration of a chemical signal over time. In practice, the possible perturbations may be subject to various constraints due to technical limitations or toxicity to the cells.

\begin{figure}
	\captionsetup{width=\linewidth,skip=4pt}
	\includegraphics[width =\linewidth]{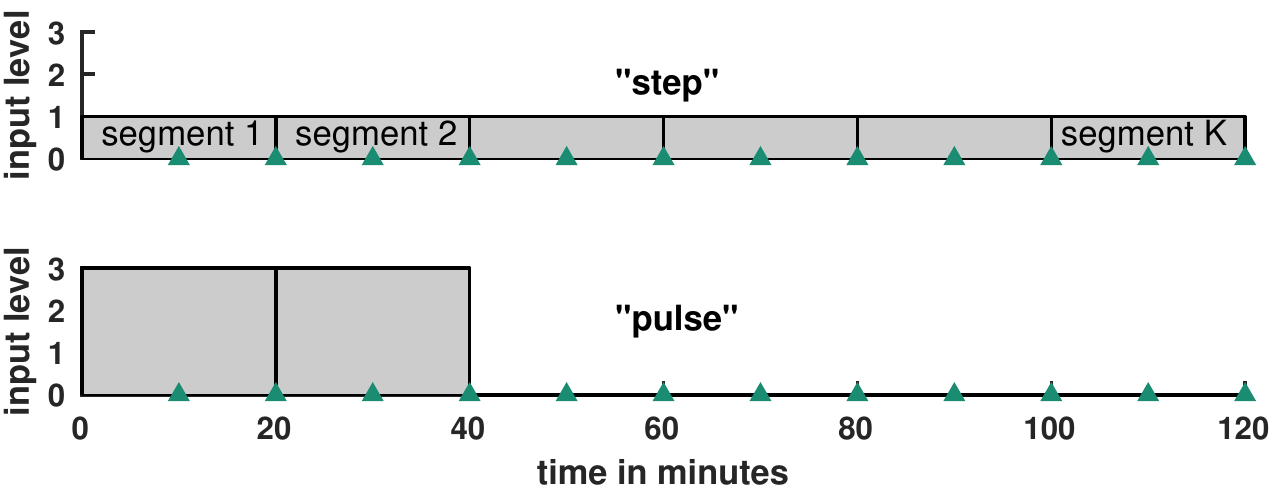}
	\caption{Schematic representation of typical 'step' and 'pulse' perturbations, the triangles indicate reporter measurement timepoints; the input sequence consists of $K$ segments of given duration (in this case 20 $\,$min). Note, both depicted perturbations have $||\mathbf{u}||_1 = 6$.}
	\label{img:input_schem}
\end{figure}

In this study, perturbations are limited to piece-wise constant functions with fixed interval length and constant $\ell_1$-norm. The time interval of the experiment $[0,T]$ is divided into $K$ segments. We denote with $u_k$ the input level at the $k$-th sub-interval (Fig. \ref{img:input_schem}). The design space is parametrized as
\begin{equation*}
\mathcal U = \left\{ (u_1, \ldots, u_K)^T \in \mathbb R_+^K : \sum_{k=1}^K u_k = E  \right\}
\end{equation*}
where $E>0$ is a constant reflecting the total dose of the input signal. We denote the discretized perturbation as a vector $\mathbf u$. 

\vspace{-2mm}

\subsection{Sampling the Design Space}

The restricted design space $\mathcal U$ still contains uncountably many perturbations. To explore $\mathcal U$, we propose a sampling algorithm based on the Metropolis-Hastings method. 

We realize this by introducing an artificial probability density
\begin{equation}  \label{eqn:p_of_u}
p ( \mathbf u ) \propto \exp \left( \beta I_{\mathbf u} ( \mathbf C, \mathbf Y )  \right)
\end{equation}
on the design space where $\beta>0$ is a design parameter.
The Metropolis-Hastings algorithm is a general method to obtain samples from distributions with unknown normalization constants \cite{Liu2001}. Given the current design parameter $\mathbf u^{(t)} $, we generate a trial value $\mathbf u'$ from a suitable proposal distribution $q (\mathbf u' \mid \mathbf u^{(t)} )$. The proposed move is accepted with probability $\alpha$ given by 
\begin{equation} \label{eqn:acceptance_ratio}
\alpha =\min \left\{ 1 ,  \frac{p(\mathbf u')}{p(\mathbf u^{(t)} ) } \frac{ q (\mathbf u^{(t)} \mid \mathbf u' ) }{ q (\mathbf u' \mid \mathbf u^{(t)} ) } \right\} \, .
\end{equation}
For a fixed number of input segments $K$, the design space $\mathcal U$ corresponds to the set of positive piece-wise constant functions with fixed step size and constant norm. This is equivalent to the standard $K$-dimensional probability simplex. A suitable proposal distribution on this space is given by a Dirichlet distribution (\cite{Kotz2000}) of the form
\begin{equation*}
q (\mathbf u' \mid \mathbf u^{(t)} ) = \mathrm{DIR} ( \mathbf u' \mid b \mathbf u^{(t)} )
\end{equation*}
where $b$ is a tuning constant controlling the variance of the proposal. With this proposal and $p(\mathbf u)$ as in Eqn. \ref{eqn:p_of_u}, the acceptance ratio $\alpha$ can be evaluated analytically. The resulting formula is straightforward but lengthy and thus not presented here. 
	\vspace{-4mm}
\begin{figure}[h]
	\captionsetup{width=\linewidth,skip=4pt}
	\includegraphics[width = \linewidth]{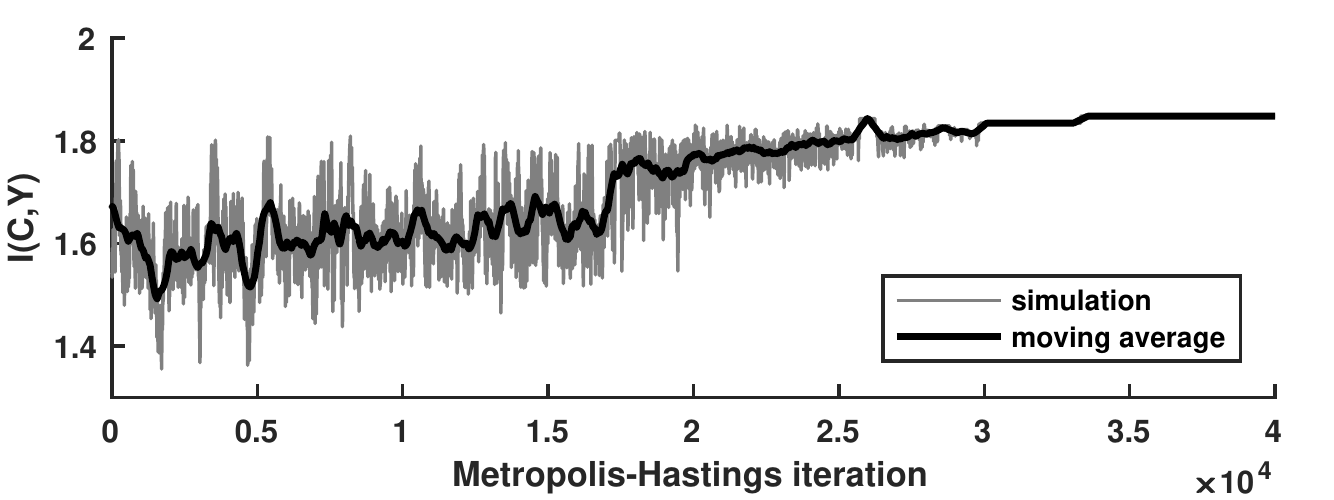}
	\caption{Example run of the Metropolis-Hastings sampler with simulated annealing showing the convergence to an approximate optimum of the mutual information. The thick line indicates a moving average with window size 500. }
	\label{fig:annealFreeze}
\end{figure}

In an optimal design scenario, we are primarily interested in the modes of $p(\mathbf u)$ because by construction the maximum of $p(\mathbf u)$ coincides with the maximum of $I_{\mathbf u}$. Hence, simulated annealing (\cite{Liu2001}) is employed by recursively increasing $\beta$ (Eqn. \ref{eqn:p_of_u}) during the simulation to achieve the result shown in Fig.  \ref{fig:annealFreeze}. The annealing shifts probability mass from flat regions near the tails towards the mode of the distribution until convergence to an approximate optimum).  


\section{Results}
\label{sec:results}
We apply the proposed method to the gene expression model described in Section \ref{sec:chemical_kinetics} with the parameter configuration in Table \ref{tab:modelA} (gene copy number of one). We consider not only the proposed best perturbation sequences, but also the worst, to gain some intuitive insight into how these come to be. First, the reference case results are presented, followed by the same model with a 'faster' reporter. The effect of the  $\ell_1$-norm on the approximately optimal input vectors $\mathbf{u^*}$ is analysed. Finally, we analyse the effect of changing the reporter kinetics for the characteristic best found perturbation sequences. 
\vspace{-1mm}
\begin{table}[h]
	\captionsetup{width=\linewidth,skip=4pt}
	\caption{Reference parameter configuration for the transcription model and coefficient of variation (CV) for priors.}
	\label{tab:modelA}
	\begin{tabularx}{\linewidth}{lcccccccc}
		\toprule
		Param.& $c_1$ & $c_2$ & $c_3$ & $c_4$ & $c_5$ & $c_6$ & $c_7$ &  \\ 
		\midrule
		Mean$/$\si{\per \second} & $0.006$ & $0.005$ & $1$ & $0.02$ & $0.01$ & $0.01$ & $0.0004$  \\ 
		CV & $10^{-4}$ &  $0.5$ & $0.005$ & $0.5$ & $0.5$& $10^{-4}$ & $10^{-4}$ \\ 
		\bottomrule
	\end{tabularx}
\end{table}
\vspace{-3mm}

We assume prior knowledge of the rate constants for all simulations (for example from a previous inference experiment). The Gamma distributed prior is given by 
\begin{equation*}
p(\mathbf{c}) = \prod_{i=1}^\nu \Gamma(a_i,b_i)
\end{equation*}
with $a_i = \mathrm{CV}_i^{-2}$ and $b_i = c_i \cdot \mathrm{CV}_i^2$;  $c_i$ and the corresponding coefficient of variation $\mathrm{CV}_i$ from the parameter configuration given in Table \ref{tab:modelA} (unless otherwise stated). 

The input is temporally discretized into $K=6$ segments of constant input levels, each of $\SI{20}{\min}$ duration. The $\ell_1$-norm is held constant for each optimization scenario. Simulated measurements of the reporter abundance corrupted by independent log-normal observation noise are recorded at intervals of $\SI{10}{\min}$, in line with common experimental protocols (for example \cite{Schneider2017}). Input sequences are optimized with regard to the inference of parameters $c_2$ to $c_5$.


The joint log-normal model imposes strong assumptions on the distribution $p(\mathbf c, \mathbf y)$ because it implicitly enforces a linear relationship in the log domain between the parameters and the measurement for a fixed perturbation $\mathbf{u}$. 

Due to the complexity of the model, it is challenging to assess the quality of such an approximation rigorously. As a simple sanity check, we considered histograms of marginal measurement distribution $p(y_i)$ at individual time-steps and scatter plots of the pairwise marginals $p(c_j, y_i)$ of the joint distribution. These are deemed to follow the log-normal distribution well. 

A more rigorous validation of the proposed method, for example by quantitative comparison with the earlier approaches, is left for future work, however, the characteristics of the optimal perturbations found with our method are in agreement with those reported  in \cite{Zechner2012a}.

\subsection{Input Optimization Results}

\begin{figure}[h]
	\captionsetup{width=1\linewidth,skip=2pt}
	\includegraphics[width =1\linewidth]{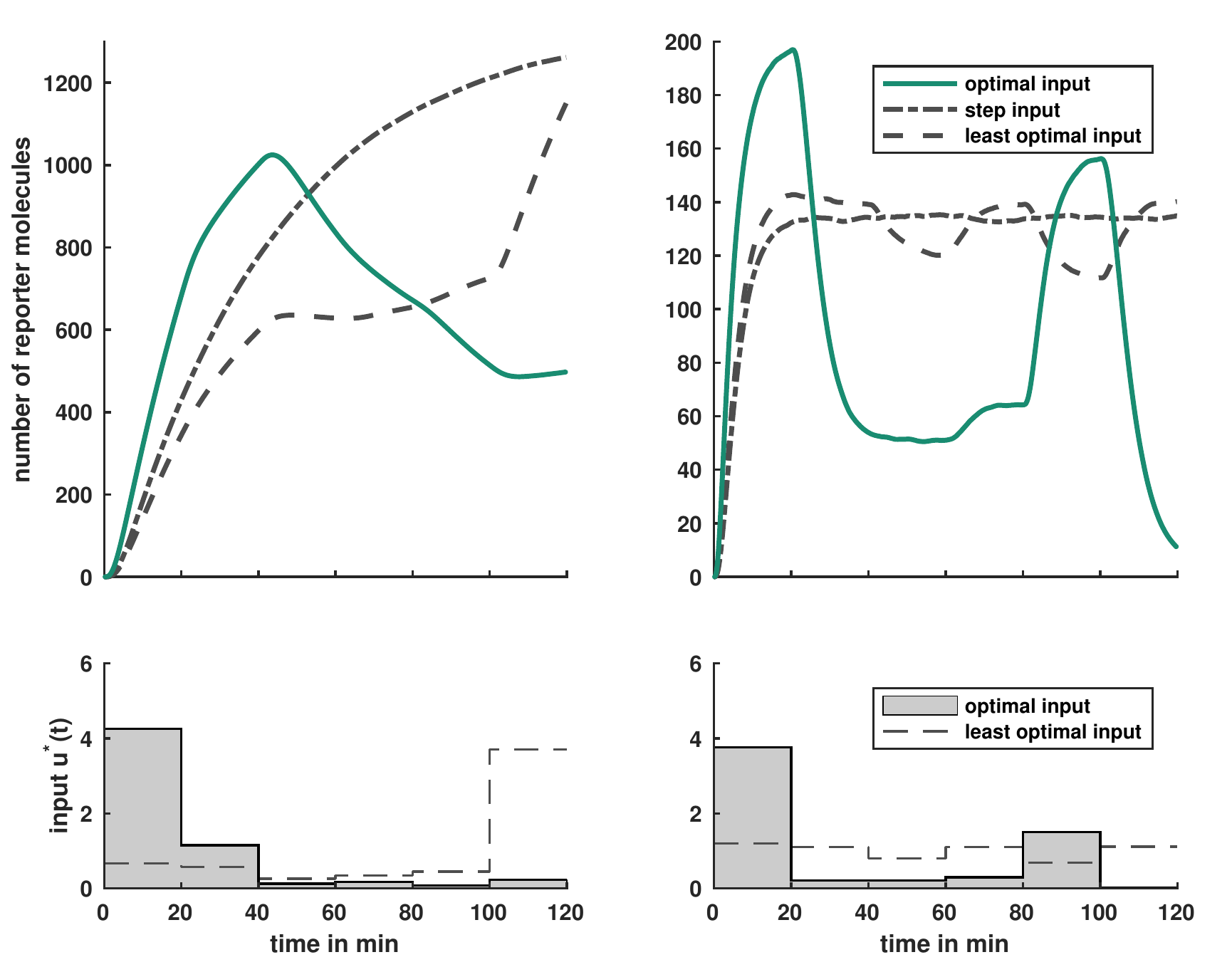}
	\caption{Mean reporter abundance (top) for the optimized perturbation sequence $\mathbf{u^*}$ (bottom); reference (left) and (right) model with larger reporter maturation and decay rates ($c_6 = \SI{0.1}{\per \second}, c_7 = \SI{0.004}{\per \second}$).}
	\label{fig:result_AC}
\end{figure}

We first consider optimizing the input for the reference model, with a reporter that matures somewhat more quickly than the fastest known fluorescent proteins ($1 / c_6 \leq T_m$, expected maturation time $T_m \approx \SI{5}{\min}$ \cite{Milo2015}). For this case, the most and least optimal perturbations are depicted in Fig. \ref{fig:result_AC} (left). 
	\vspace{-4mm}
		
The best sequence is an early pulse, while a late pulse is found to coerce the least informative behaviour of the cells. The mean reporter trajectory in the optimal case exhibits dynamics that include both an accumulation and a decay phase, whereas the step perturbation and the least optimal variant merely capture reporter accumulation and omit the decay phase. These findings are in agreement with results reported in \cite{Zechner2012a}. The characteristics of the optimal and least optimal perturbation sequences found are similar for slower reporters with maturation times of up to $\SI{45}{\min}$, corresponding to typical \textit{in-vivo} reporters in yeast. 

'Faster' reporters are available, which follow a biomolecuar systems dynamics without maturation delays. Translocation reporters shuttle mature fluorescent proteins between the nucleus and cytoplasm of a cell within $\SI{1}{\min}$ \cite{Aymoz2016}. We simulate their behaviour by increasing $c_6$ and $c_7$ associated with reporter maturation and decay. 

The characteristic optimal and least optimal inputs for a faster reporter are shown in Fig. \ref{fig:result_AC} (right) with the corresponding mean reporter trajectories. The faster reporter enables two accumulation-decay cycles to be observed within the allotted measurement duration. The optimal perturbation is a double pulse, with an early and a later pulse. The two reporter accumulation-decay cycles are positioned as far apart as possible such that the reporter nears extinction at the final measurement timepoint. 

In contrast to the result for the reference model, the least optimal perturbation for the faster reporter is not a late pulse but similar to the standard step perturbation with slight fluctuations. The protein traces for both the step and the late pulse perturbations achieve a steady state by the second measurement timepoint and remain there throughout the experiment.

\subsection{Variation of Mean Perturbation Strength}
\begin{figure}[h]
	\captionsetup{width=\linewidth,skip=2pt}
	\includegraphics[width =\linewidth]{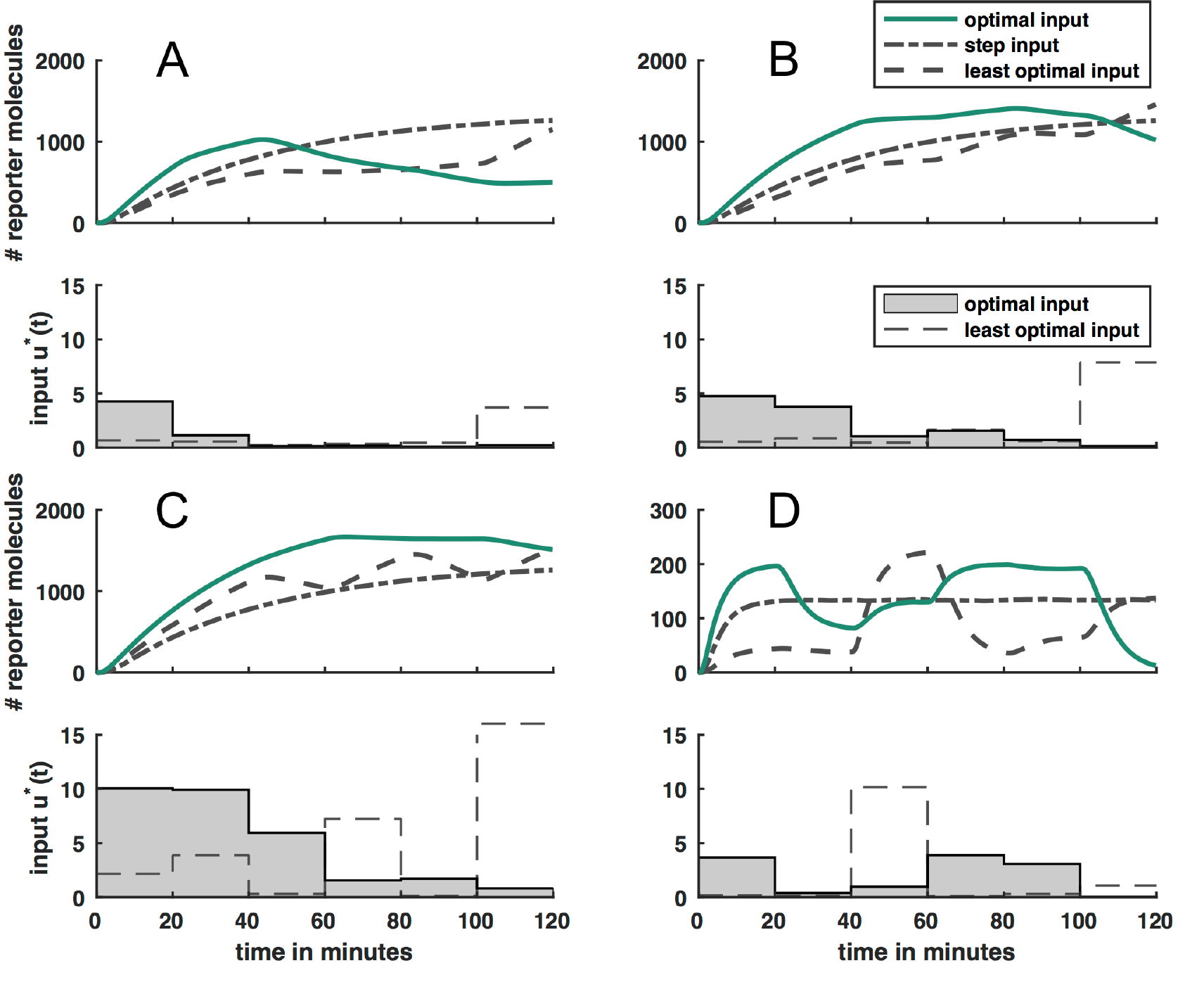}
	\caption{Optimal input sequences and reporter trajectories with various $\ell_1$-norms ($||\mathbf{u}||_1/K = 1,2$ and $5$, A to C respectively) for the reference model, and for the faster reporter model $||\mathbf{u}||_1/K=2$ (D).}
	\label{fig:result_L1}
		\vspace{-2mm}
\end{figure}

The effect of varying the $\ell_1$-norm of the input sequence on the character of optimal and least optimal perturbations was investigated. The results are depicted in Fig. \ref{fig:result_L1}. No substantial effect was found for the standard reporter, beyond a widening of the early pulse (Fig. \ref{fig:result_L1} A to C). At $||\mathbf{u}||_1 / K = 5$ (Fig. \ref{fig:result_L1} C) the least optimal perturbation is a series of pulses of increasing strength, with the majority of the input being contained in the final pulse at the latest timepoint. 
	\vspace{-2mm}

The character of the optimal perturbation is also conserved for stronger stimulation of the faster reporter (between Fig. \ref{fig:result_AC} (right) and Fig. \ref{fig:result_L1} D). An early pulse is followed by a decay phase and a late, somewhat wider, pulse, which absorbs most of the additional input strength. The least optimal perturbation, however, is markedly different from the step found for $||\mathbf{u}||_1 / K = 1$ (Fig. \ref{fig:result_AC}). Almost the complete weight of the perturbation is concentrated into a single pulse at the middle of the experiment. The resulting trajectories are near extinction apart from a short peak after the stimulation. 

The least optimal perturbations are more dependent on the $\ell_1$-norm for both models. In both cases the poor input sequences achieve higher peaks than the optimal ones. For high peaks in stimulation the gene is rarely turned off and therefore strong modulation of the gene-on rate $c_1$ becomes irrelevant as the propensity is zero in the gene-on state. 

\subsection{Analysis of Reporter Rates for Characteristic Best Input}
The optimization of the input perturbations in the previous sections found two characteristic sequences: the 'early pulse' and the 'double pulse'. To gain an intuitive understanding of when to use which sequence and how experimental constraints imposed by the reporters influence the inference capacity, we now analyse the effect of reporter kinetics on the mutual information $I(\mathbf{C},\mathbf{Y})$. The ratio of reporter maturation and decay rates ($c_7/c_6 = \mathrm{const.}$) is held constant. The reporters considered in the previous sections have ratio of $c_6/c_5 = 1$ and $c_6/c_5 = 10$ respectively. 
	\vspace{-4mm}
\begin{figure}[h]
	\captionsetup{width=\linewidth,skip=4pt}
	\includegraphics[width = \linewidth]{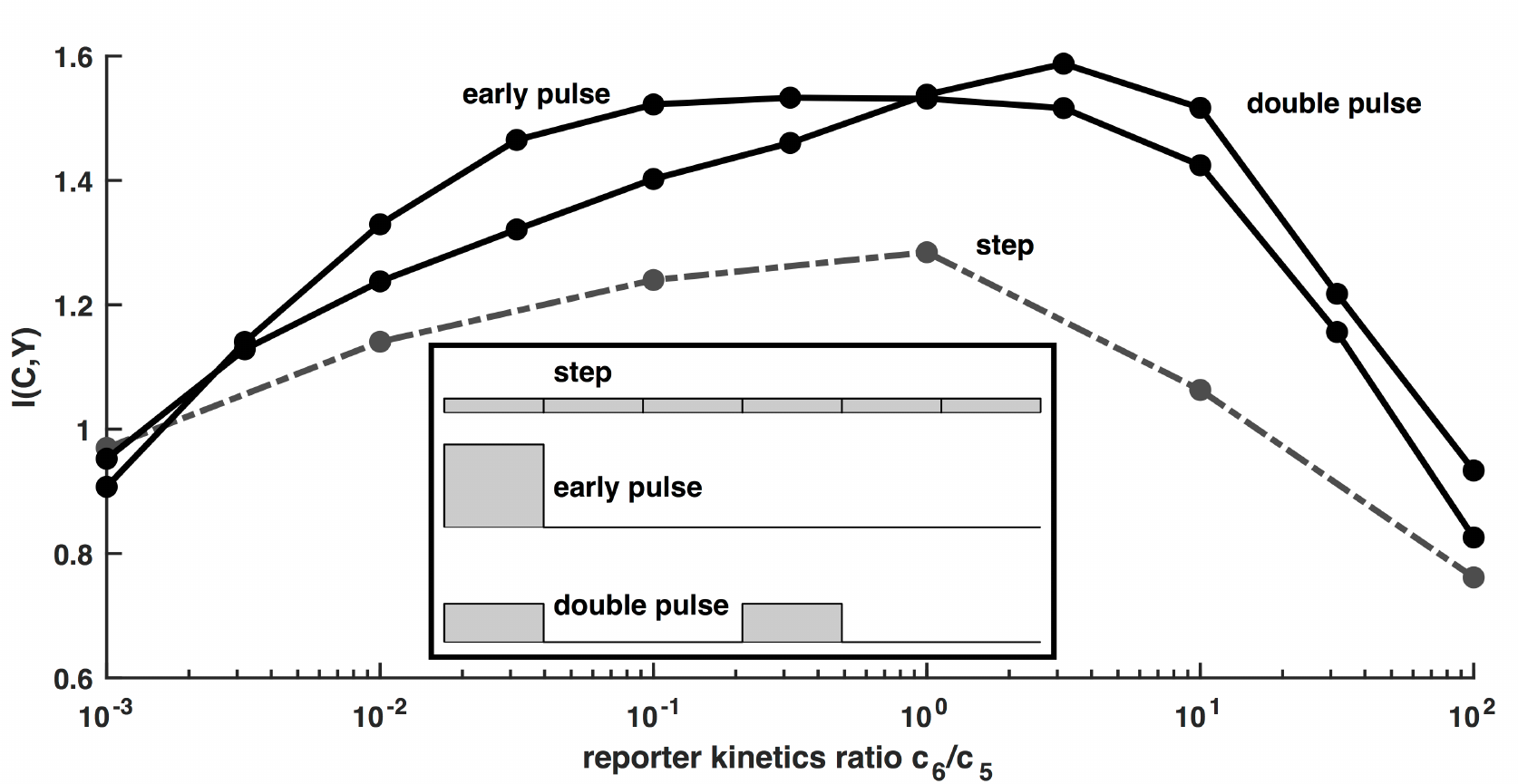}
	\caption{Mutual information $I(\mathbf{C},\mathbf{Y})$ of characteristic perturbations over the ratio of reporter production and maturation rates ($c_6/c_5$) for a constant ratio of maturation and decay rates; large ratio denote nominally 'faster' reporters and all perturbations satisfy $||\mathbf{u}||_1 = 6$.}
	\label{fig:reporterSpeed}
		\vspace{-2mm}
\end{figure}
The mutual information for the two characteristic perturbation profiles over various ratios of reporter production and maturation rates ($c_6/c_5$) are shown in Fig. \ref{fig:reporterSpeed}. The results indicate that for slower reporters the early pulse is most informative. For faster reporters, however, the double pulse is the most informative. 

At the extreme ends of the reporter speed scale, both very slow and very fast reporters were found to deliver lower information gains, seemingly regardless of perturbation strategy. The very slow reporters are hindered by the time constraint $T=\SI{120}{\min}$, while faster reporters tend to decay toward extinction quickly.

These results indicate that it may generally be beneficial to include as many reporter production-decay cycles into an experiment as possible. However, optimization of the input perturbations is expected to lead to the most informative experimental results. The proposed method for maximizing the information gain for can both determine whether to employ a single or a double pulse, or optimize the input perturbation for a given reporter.  


\section{Conclusion}
We propose an approximate method to perform optimal experimental design of chemical perturbation profiles for the inference of biomolecular circuit parameters fields of synthetic biology and molecular communication. Our method scans the entire input parameter space, finding both the most and least optimal input sequences. This may be beneficial to gain deeper insights into what constitutes a good perturbation for the continuous-time Markov chains considered, as well as information on how intermediate profiles perform and how wide the optimal peaks are. 

An advantage of our method is that it does not require an inference step. Forward simulation of the process suffices to determine the approximately optimal perturbation designs. Furthermore, the Markov chain Monte Carlo framework employed can easily be extended to other scenarios and optimize other experimental parameters such as the timing of measurements, for example. 

While we are cautious with regard to interpreting the results due to the complexity of the system investigated, the results attained for two test cases showed two distinct characteristic perturbations. These were found to depend on the properties of the employed reporters (for a given system). Relatively fast reporters benefited from repeated stimulation, while slower reporters achieved the most informative results for a strong early perturbation. In the future, our tool may be employed to generalize the characteristics of 'optimal' perturbations as the basis for intuitive design rules. 

\begin{acks}
This work is supported by LOEWE CompuGene.
\end{acks}

\bibliographystyle{ACM-Reference-Format}
\bibliography{NanoCom2018Bib}

\clearpage
\appendix

\clearpage

\end{document}